\newif\iffinal
\finalfalse	
\finaltrue 

\newif\ifmarek
\marektrue

\documentclass[reqno,twoside,11pt]{amsart}
\iffinal\else\usepackage[notref,notcite]{showkeys}\fi
\usepackage{cite}
\usepackage{subcaption}
\usepackage{amsmath}
\usepackage{amsfonts}
\usepackage{amssymb}
\usepackage{graphicx}

\usepackage{graphics}
\graphicspath{{/Users/aubain2015/Documents/york/}}
\usepackage{verbatim}
\usepackage{color}
\usepackage{float}
\usepackage{dirtytalk}
\IfFileExists{epsf.def}{\input epsf.def}{\usepackage{epsf}}

\ifmarek
\IfFileExists{myowntimes.sty}{\usepackage{myowntimes}\usepackage{mathrsfs}}
	{\usepackage{times}\newcommand{\mathscr}{\mathcal}}

\usepackage{mathptmx}
\fi

\DeclareFontFamily{OT1}{eusb}{} \DeclareFontShape{OT1}{eusb}{m}{n}
{<5> <6> <7> <8> <9> <10> <11> <12> <14.4> eusb10}{}
\DeclareMathAlphabet{\eusb}{OT1}{eusb}{m}{n}

\DeclareFontFamily{OT1}{eusm}{} \DeclareFontShape{OT1}{eusm}{m}{n}
{<5> <6> <7> <8> <9> <10> <11> <12> <14.4> eusm10}{}
\DeclareMathAlphabet{\eusm}{OT1}{eusm}{m}{n}

\DeclareFontFamily{OT1}{eufm}{} \DeclareFontShape{OT1}{eufm}{m}{n}
{<5> <6> <7> <8> <9> <10> <11> <12> <14.4> eufm10}{}
\DeclareMathAlphabet{\mathfrak}{OT1}{eufm}{m}{n}

\DeclareFontFamily{OT1}{fraktura}{}
\DeclareFontShape{OT1}{fraktura}{m}{n} {<5> <6> <7> <8> <9> <10>
<11> <12> <13> <14.4> [1.1] eufm10}{}
\DeclareMathAlphabet{\fraktura}{OT1}{fraktura}{m}{n}

\DeclareFontFamily{OT1}{cmfi}{} \DeclareFontShape{OT1}{cmfi}{m}{n}
{<5> <6> <7> <8> <9> <10> <11> <12> <13> <14.4> [0.9] cmfi10}{}
\DeclareMathAlphabet{\cmfi}{OT1}{cmfi}{b}{n}

\DeclareFontFamily{OT1}{cmss}{} \DeclareFontShape{OT1}{cmss}{m}{n}
{<5> <6> <7> <8> <9> <10> <11> <12> <13> <14.4> cmss10}{}
\DeclareMathAlphabet{\cmss}{OT1}{cmss}{m}{n}

\newtheoremstyle{thm}{1.5ex}{1.5ex}{\itshape\rmfamily}{}
{\bfseries\rmfamily}{}{2ex}{}

\newtheoremstyle{def}{1.5ex}{1.5ex}{\slshape\rmfamily}{}
{\bfseries\rmfamily}{}{2ex}{}

\newtheoremstyle{rem}{1.3ex}{1.3ex}{\rmfamily}{}
{\itshape}
{} {1.5ex}{}

\theoremstyle{thm}
\newtheorem{theorem}{Theorem}[section]
\newtheorem{lemma}[theorem]{Lemma}

\theoremstyle{def}

\theoremstyle{rem}

\numberwithin{equation}{section}

\renewcommand{\section}{\secdef\sct\sect}
\newcommand{\sct}[2][default]{\refstepcounter{section}
\addcontentsline{toc}{section}
{{\tocsection {}{\thesection}{\!\!\!\!#1\dotfill}}{}}
\vspace{0.7cm}
\centerline{ 
\scshape\arabic{section}.\ #1} \nopagebreak \vspace{0.2cm}}
\newcommand{\sect}[1]{
\vspace{0.4cm} \centerline{\large\scshape\rmfamily #1}
\vspace{0.2cm}}

\renewcommand{\subsection}{\secdef\subsct\sbsect}
\newcommand{\subsct}[2][default]{\refstepcounter{subsection}
\addcontentsline{toc}{subsection}
{{\tocsection{\!\!}{\hspace{1.2em}\thesubsection}{\!\!\!\!#1\dotfill}}{}}
\nopagebreak\vspace{0.45\baselineskip} {\flushleft\bf
\thesection.\arabic{subsection}~\bf #1.~}
\\*[3mm]\noindent
\nopagebreak}
\newcommand{\sbsect}[1]{\vspace{0.1cm}\noindent
\textbf{#1.~}\vspace{0.1cm}}

\renewcommand{\subsubsection}{%
\secdef \subsubsect\sbsbsect}
\newcommand{\subsubsect}[2][default]{%
\refstepcounter{subsubsection} 
\addcontentsline{toc}{subsubsection}{{\tocsection{\!\!}
{\hspace{3.05em}\thesubsubsection}{\!\!\!\!#1\dotfill}}{}}
\nopagebreak
\vspace{0.15\baselineskip} \nopagebreak {\flushleft\rmfamily
\itshape\arabic{section}.\arabic{subsection}.\arabic{subsubsection}
\ \rmfamily #1\/.}\ }
\newcommand{\sbsbsect}[1]{\vspace{0.1cm}\noindent
\rmfamily \itshape
\arabic{section}.\arabic{subsection}.\arabic{subsubsection} \
\sffamily #1\/.\ }

\iffinal

\else

\fi

\newcommand{\scrF}{\mathscr{F}}

\title[]
{\large Epidemic Dynamics and adaptive vaccination strategy:\\renewal equation  approach}

\author[]{Aubain Nzokem$^1$, Neal Madras$^2$}
\thanks{$^1$ aubain14@my.yorku.ca}
\thanks{$^2$ madras@yorku.ca}
\begin{document}

\maketitle

\vspace{-2mm}
\centerline{\textit{Department of Mathematics \& Statistics, York University, Toronto}}

\vspace{-2mm}
\begin{abstract}
We use analytical methods to investigate a continuous vaccination strategy's effects on the infectious disease dynamics  in a closed  population  and a demographically open population.  The methodology and  key assumptions are based on Breda et al (2012).
We show  that the cumulative force of infection  for the closed population and the endemic force of infection in the demographically open population can be reduced significantly  by combining  two factors:  the vaccine effectiveness  and the vaccination rate. The impact of these factors on the force of infection  can transform an endemic steady state into a disease-free state. \\
\keywords{ Keywords: Force of infection, Cumulative force of infection, Scalar-renewal equation, Per capita death rate, Lambert function, adaptive vaccination strategy}
 \end{abstract}
\section{Introduction}
\noindent
The paper of Kermack and McKendrick (1927)  is one of the best known  contributions to the mathematical  theory of epidemic modelling. The paper  provides the condition of outbreak  and the final size equation in a closed population setting. One of the key features of Kermack and McKendrick (1927) was to introduce an age of infection model. In such a model, the general  infectivity function ($A(\tau)$) of an individual  is considered and depends on the time ($\tau$) elapsed since the infection took place.   Kermack and McKendrick's framework encompasses a wide family of epidemic models; Breda et al (2012) have illustrated the generalisation by providing the following age infection functions for standard SIR and SEIR models.
\begin{equation} 
\label{eq:1}
\begin{split}
A(\tau)&=\beta e^{-\alpha \tau} \Longleftrightarrow  SIR \\
A(\tau)&=\beta \frac{\gamma}{\gamma - \alpha}(e^{-\alpha \tau} -e^{-\gamma \tau})\Longleftrightarrow SEIR\\ 
\end{split}
\end{equation}
\noindent
The paper of Breda et al (2012) \say{On the formulation of epidemic models (an appraisal of Kermack and McKendrick)} revised Kermack and McKendrick's  paper and produced the same results, but the method used was different. In fact, Breda et al (2012) considered  the force of infection as a result of a nonlinear scalar-renewal equation, and they analyzed the cumulative force of infection or the simple force of infection at the disease-free equilibrium and  the endemic equilibrium.\\
\noindent
 In the current paper, we  investigate the effects of an adaptive vaccination strategy on the dynamics of infectious diseases in a closed  population  and a demographically open population. The methodology and  key assumptions are based on Breda et al (2012). \\
\begin{figure}[h]
\vspace{-20pt}
\centering
\includegraphics[width=0.5\textwidth]{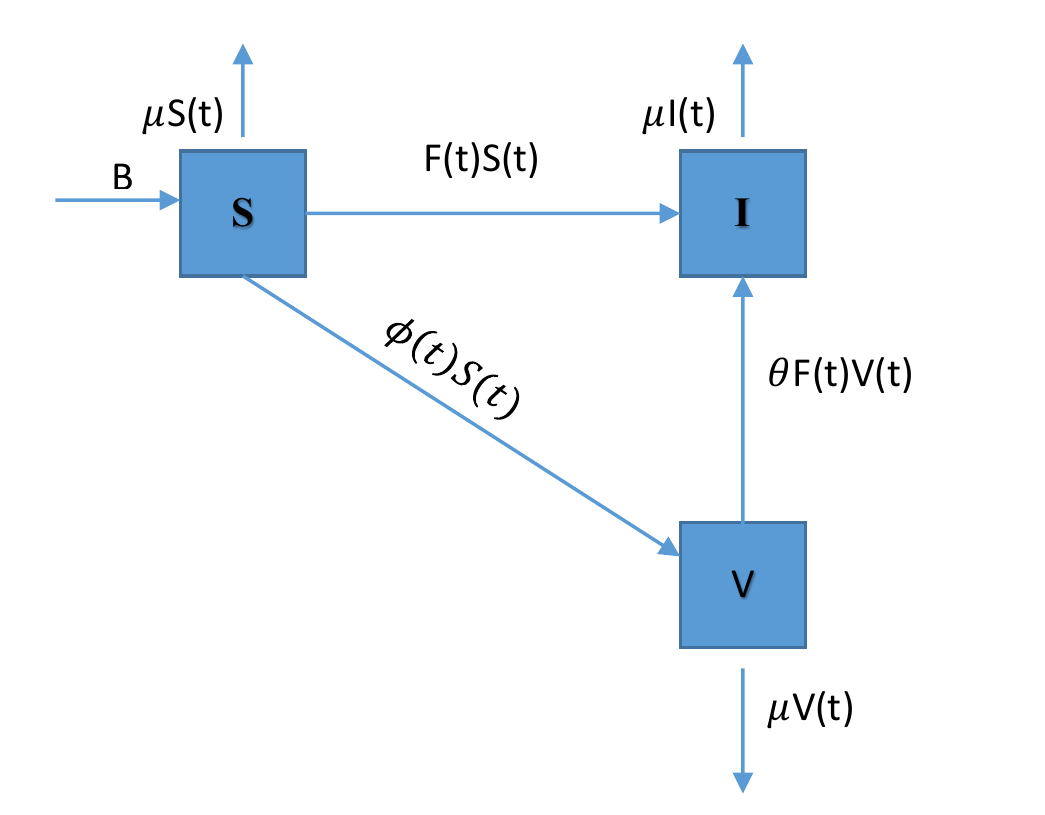} 
\vspace{-15pt}
\caption{
The transfer diagram of the model: the force of infection function (the probability per unit of time that a susceptible becomes infected)   $t \mapsto F(t)$; the rate of vaccination function $t\mapsto\phi(t)$; $S$ 
(non-vaccinated susceptible); $V$ (vaccinated susceptible); $\theta$ (vaccine parameter ($0\leq\theta\leq1$)); $I$ (infected population); $B$ (constant birth rate); $\mu$ (constant per capita death rate).}
\vspace{-10pt}
\label{fig:diag}
\end{figure} 

\noindent
The epidemic model and the vaccination process are illustrated by Figure \ref{fig:diag}. The susceptible population is  divided into non-vaccinated susceptible and vaccinated susceptible. 
The main assumptions are that  the infection leads to permanent immunity (no re-infection); the force of infection occurring in non-vaccinated susceptibles  is proportional to the force of infection  within the vaccinated susceptibles; and the rate of vaccination is proportional to the force of infection in the non-vaccinated susceptibles. In fact, a natural feature of an adaptive vaccination policy is that the rate of vaccination should increase when the force of infection increases and decrease when the force of infection decreases. 
Our paper will first analyze the closed population setting, and then introduce a survival function and analyze  an age-structured population.

\section{Closed population epidemic model}
\noindent
The dynamic of infection can be described in each susceptible group as illustrated in the transfer diagram of the model.  The instantaneous change in the susceptibles is determined by the number of new cases of susceptibles infected per unit of time (incidence) and the number of new vaccinated susceptibles per unit of time. In a closed population, the number of susceptibles only changes due to transmission of infection and vaccination; thus, $\mu=B=0$ in Figure \ref{fig:diag}. 
The following system of equations can be derived.
\begin{equation} 
\label{eq:2}
\begin{split}
\frac{dS(t)}{dt}&= -F(t)S(t)-\phi(t)S(t) ,\\
\frac{dV(t)}{dt}&= -\theta F(t)V(t)+\phi(t)S(t) . 
\end{split}
\end{equation}
\noindent
The initial conditions are given by $S(-\infty)>0$ and $V(-\infty)=0$.\\
\noindent
By solving the system of equations (\ref{eq:2}), we have the  following results.
\begin{equation} 
\label{eq:6}
\begin{split}
S(t)&\;=\;S(-\infty)e^{-(y+\Phi)(t)} , \\
(S+V)(t)&\;=\;S(-\infty)C(t)e^{-\theta y(t)} ,\\
\end{split}
\end{equation}
where
\begin{equation} 
\label{eq:6a}
\begin{split}
 y(t)&\;=\; \int_{-\infty}^{t}F(\sigma)d\sigma,  \quad \Phi(t)\;=\;\int_{-\infty}^{t}\phi(\sigma)d\sigma \,,
  \quad  \hbox{and}
  \\ 
 C(t)&\;=\; 1-(1-\theta)\int_{-\infty}^{t}F(\sigma)e^{-(1-\theta)y(\sigma)-\Phi(\sigma)}d\sigma \,.  \\
\end{split}
\end{equation}

\noindent
The force of infection depends on the size of the infectious population. The rate of new
infections at time $t-\tau$ is $F(t-\tau)S(t-\tau)$ from the non-vaccinated susceptibles and 
$\theta F(t-\tau)V(t-\tau)$ from the vaccinated susceptibles. After  $\tau$ additional units of time, these
cases contribute  
$(F(t-\tau)S(t-\tau)+ \theta F(t-\tau)V(t-\tau))A(\tau)$ to the force of infection at time $t$. 
By summing all the contributions with respect to the elapsed time $\tau$, we obtain 
the scalar-renewal equation
\begin{equation} 
\label{eq:3}
F(t)=\int_{0}^{\infty} (F(t-\tau)S(t-\tau)+ \theta F(t-\tau)V(t-\tau))\,A(\tau)\, d\tau .
\end{equation}
\noindent
Taking into account the equations (\ref{eq:3}) and (\ref{eq:2}), we derive the following cumulative force of infection at each time $t$, denoted $y(t)$.
\begin{equation} 
\label{eq:4}
\begin{split}
y(t) & =\int_{-\infty}^{t}F(\sigma)d\sigma \\
    &= \int_{0}^{\infty}\int_{-\infty}^{t}-\frac{d(S+V)}{dt}(\sigma-\tau)\,\,A(\tau)\,d\tau \,d\sigma \\
     & =\int_{0}^{\infty} S(-\infty)\left(1-\frac{(S+V)(t-\tau)}{S(-\infty)}\right)\, A(\tau)\,d\tau .
 \end{split}
\end{equation}
\noindent
When $t \rightarrow \infty$, the equation (\ref{eq:4}) becomes
\begin{equation} 
\label{eq:5}
\begin{split}
y(\infty)&\;=\;  \left(1-\frac{(S+V)(\infty)}{S(-\infty)}\right)\int_{0}^{\infty} S(-\infty)\,A(\tau)\,d\tau \\
     & =\; R \left(1-\frac{(S+V)(\infty)}{S(-\infty)}\right) \hspace{10mm}
 \hbox{where} \hspace{5mm}
 R\;=\;\int_{0}^{\infty} S(-\infty)\,A(\tau)\,d\tau \,.
 \end{split}
\end{equation}
\noindent
In the expression for $C(t)$ in \eqref{eq:6a}, the integral  
$\int_{-\infty}^{t}F(\sigma)e^{-(1-\theta)y(\sigma)-\Phi(\sigma)}d\sigma\ $ is not easy to handle in general, 
as the rate of vaccination  ($\phi(t)$) is unknown. We consider a special case by  assuming  a linear relationship between the  vaccination rate  and the force of infection. Specifically, we assume 
$\phi(t)=pF(t)$ where $p$ is the vaccination rate parameter. \\
With this assumption on $\phi(t)$, the following results can be derived.
\begin{equation} 
\label{eq:7}
\begin{split}
C(\infty)&\;=\;\frac{1}{1+p-\theta}\left(p+(1-\theta)e^{-(1+p-\theta)y(\infty)}\right)\,,  \\
\frac{(S+V)(\infty)}{S(-\infty)} &\;=\;C(\infty)e^{-\theta y(\infty)}\\
                                 &\;=\;  \frac{1}{1+p-\theta}\left(pe^{-\theta y(\infty)}+(1-\theta)e^{-(1+p)y(\infty)}\right) \,.
\end{split}
\end{equation}

\subsection{Impact on the endemic steady state}
\noindent
The expression in \eqref{eq:7} was replaced into the equation  \eqref{eq:5}, which describes the asymptotic behaviour of the cumulative force of infection in  the epidemic dynamic.  The result is 
\begin{equation} 
\label{eq:8}
y(\infty) \;=\;R\left(1-\frac{pe^{-\theta y(\infty)}+(1-\theta)e^{-(1+p)y(\infty)}}{1+p-\theta} \right)
 \hspace{5mm} \hbox{with}\hspace{5mm} 
  R=\int_{0}^{\infty} S(-\infty)\,A(\tau)\,d\tau \,.
\end{equation}
\noindent
According to Breda et al (2012), the reproduction number ($R$) defined in \eqref{eq:8} can be interpreted as the expected number of secondary cases caused by a primary case introduced into a population with susceptible density $S(-\infty)$. The case $R>1$ is the interesting case to study if we want to investigate the effect of the vaccination rate parameter ($p$)
and the vaccine parameter ($\theta$) on the cumulative infection force ($y(\infty)$). \\
\noindent
\\
In order to analyse the equation (\ref{eq:8}) above, we have to know the classical properties of the solution of the equation $x=W(x)e^{W(x)}$, where the solution  W(x) is called the Lambert function.\\
\begin{figure}[h]
\vspace{-5pt}
\centering
\includegraphics[width=0.5\textwidth]{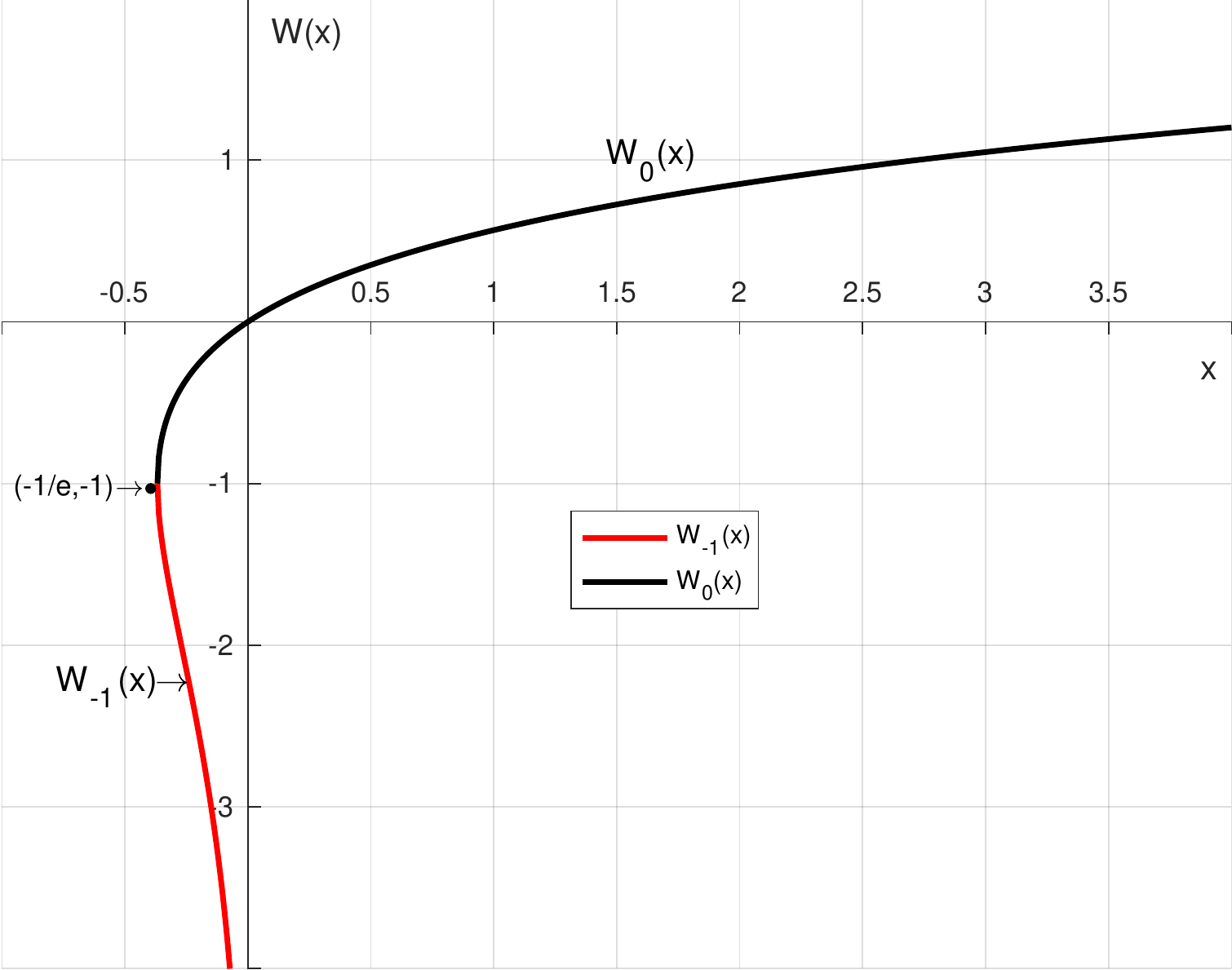} 
\vspace{-5pt}
\caption{
Solution of the equation $x=W(x)e^{W(x)}$ }
\vspace{-5pt}
\label{fig:diag2}
\end{figure} 

\noindent
As shown in the Figure \ref{fig:diag2}, the solution  $W(x)$ is actually a multivalued function, with two branches denoted $W_0$ and $W_{-1}$  defined as follows:
\begin{center} 
   \[W(x)=
  \begin{cases}
  W_0(x)   & \quad \text{if } -\frac{1}{e}\;\leq \;x \,,  \\
W_{-1}(x) & \quad \text{if } -\frac{1}{e}\;\leq \; x \;<\;0 \,.
  \end{cases}
\] 
\end{center}

\begin{lemma}
   \label{lem.W}
Let  $R$ be a positive real number, and consider nonnegative solutions $u$ of the equation
\[      u \;=\; R(1-e^{-u}) \,.    \]
(a)  If  $0<R\leq 1$, then $u=0$ is  the only nonnegative solution.
\\
(b)  If  $R>1$, then there are two nonnegative solutions:  $u=0$ and $u=R+W_0(-Re^{-R})$.
\end{lemma}
\noindent
\textbf{Proof:}\\
 Part (a): The function $g(u)=u-R(1-e^{-u})$ is strictly increasing for $0\leq u$. Therefore the unique root  is $u=0$.\\
 Part (b) : We have $u-R=-R{e^{-R}}{e^{-u+R}}$ , which implies $(u-R){e^{u-R}}=-R{e^{-R}}$. By using  the Lambert function, we have $u=R+W(-R{e^{-R}})$.\\
For $R>1$, using the identity property of the Lambert function, we have 
\begin{center} 
   \[W(-R{e^{-R}})=
  \begin{cases}
  W_0(-R{e^{-R}})  & >\; -1,  \\
W_{-1}(-R{e^{-R}}) & =\; -R.
  \end{cases}
\] 
\end{center}
\noindent
Therefore, we have  two solutions $u=0$ and $u=R+W_0(-Re^{-R})$.\\
\medskip

\noindent
\textbf{Case 1: ineffective vaccine($\theta =1$) }\\
 The expression in \eqref{eq:8} becomes
\begin{equation} 
\label{eq:9}
y(\infty) =R\left(1-e^{- y(\infty)}\right).\ \
\end{equation}
\noindent
 The quantity $y(\infty)$ is  the same as the cumulative force of infection without vaccination 
(Breda et al (2012)).  Using  Lemma \ref{lem.W}, the nonzero solution is $y(\infty)=R+W_0(-Re^{-R})$, because the hypothesis is that $R>1$.\\

\medskip 
\noindent
\textbf{Case 2: 100\% effective vaccine ($\theta=0$) } \\
The expression in \eqref{eq:8} becomes
\begin{equation} 
\label{eq:10}
y(\infty) =R\left(1-\frac{p+e^{-(1+p)y(\infty)}}{1+p}\right ).   
\end{equation}
Using  Lemma \ref{lem.W} with the substitution $u=(1+p)y(\infty)$, we find that Equation (\ref{eq:10})
has the positive solution  $y(\infty) \,=\,  [R+W_0(-Re^{-R})]/(1+p)$ when $R>1$. Thus we approach a disease-free steady state as $p$ gets large.\\

\medskip
\noindent
\textbf{Case 3: $p \rightarrow \infty$ and $\theta\neq0$}\\
 By increasing the vaccination rate parameter ($p$), the expression in \eqref{eq:8} becomes
\begin{equation} 
\label{eq:11}
y(\infty) =R\left(1-e^{-\theta y(\infty)}\right).
\end{equation}
Using  Lemma \ref{lem.W} with the substitution $u=\theta y(\infty)$, we find the solution of the equation \eqref{eq:11} to be
\[      y(\infty)\;=\;
  \begin{cases}
  0     & \quad \text{if } \theta\leq \frac{1}{R} \,, \\
R +\frac{1}{\theta}W_0(-\theta R e^{-\theta R})   
 & \quad \text{if } \frac{1}{R}< \theta\leq 1\,.
  \end{cases}
\] 
 \medskip
 
\begin{figure}[h]  
 \vspace{-6pt}
    \begin{subfigure}{.5\linewidth}
\centering
    \includegraphics[scale=0.4]{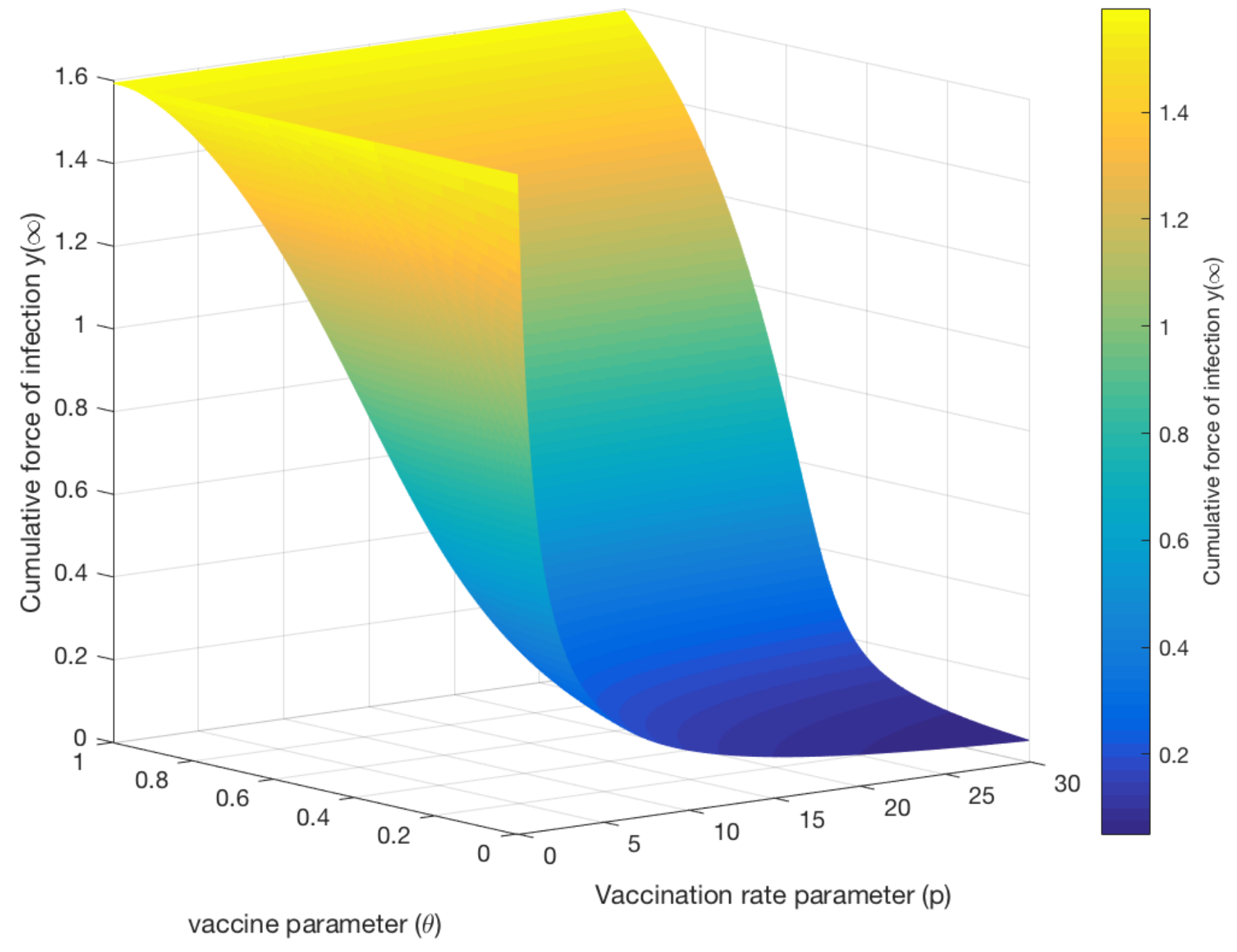}   
\caption{\tiny Cumulative force of infection $y(\infty)$ as a function of $\theta$ and $p$}
\vspace{10pt}
\label{fig:sub11}
\end{subfigure}%
\begin{subfigure}{.5\linewidth}
\centering
    \includegraphics[scale=0.4]{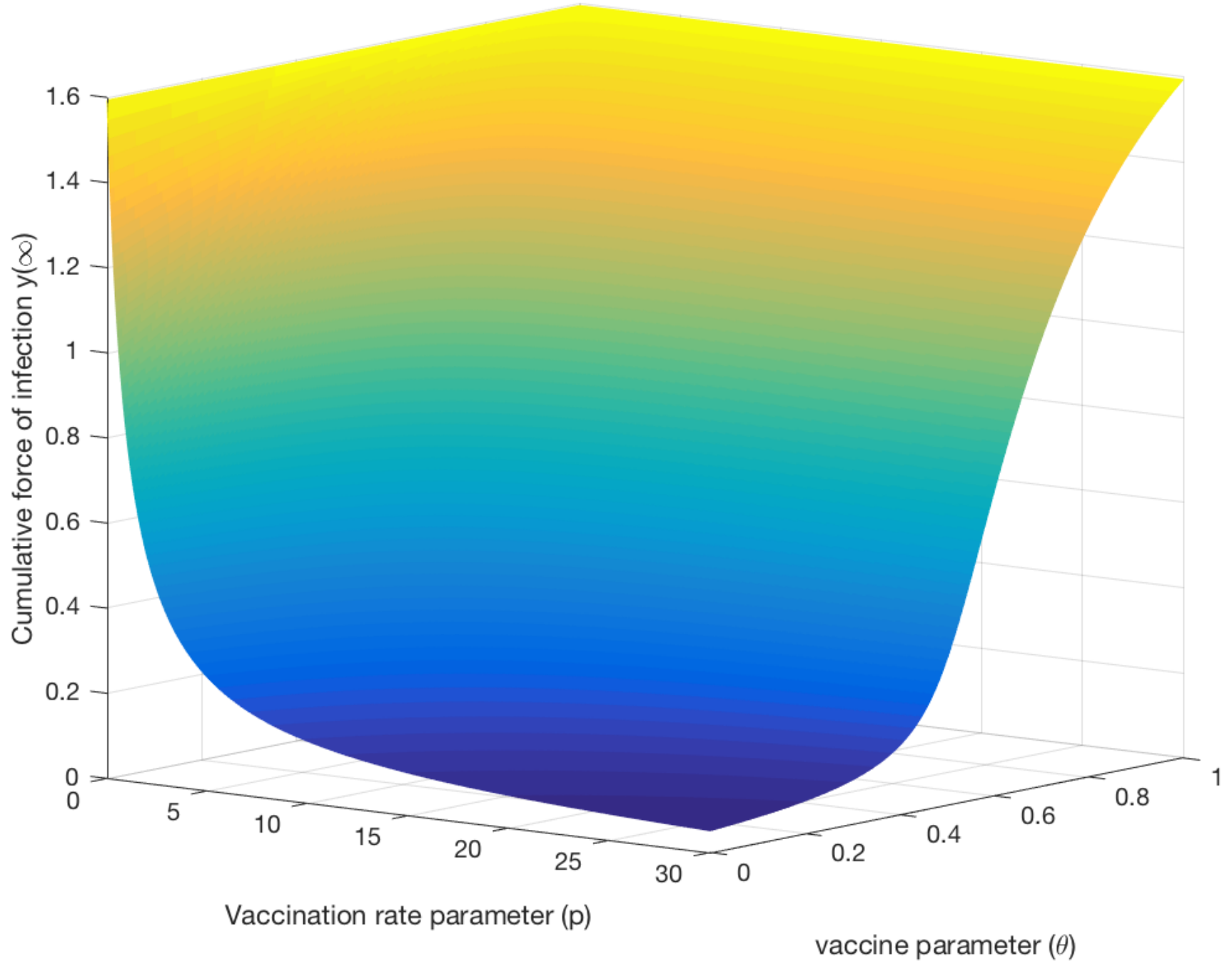}   
\caption{\tiny Cumulative force of infection $y(\infty)$ as a function of $\theta$ and $p$}
  \vspace{10pt}
\label{fig:sub12}
\end{subfigure}\\[1ex]
\begin{subfigure}{\linewidth}
\centering
 \vspace{-15pt}
    \includegraphics[scale=0.5]{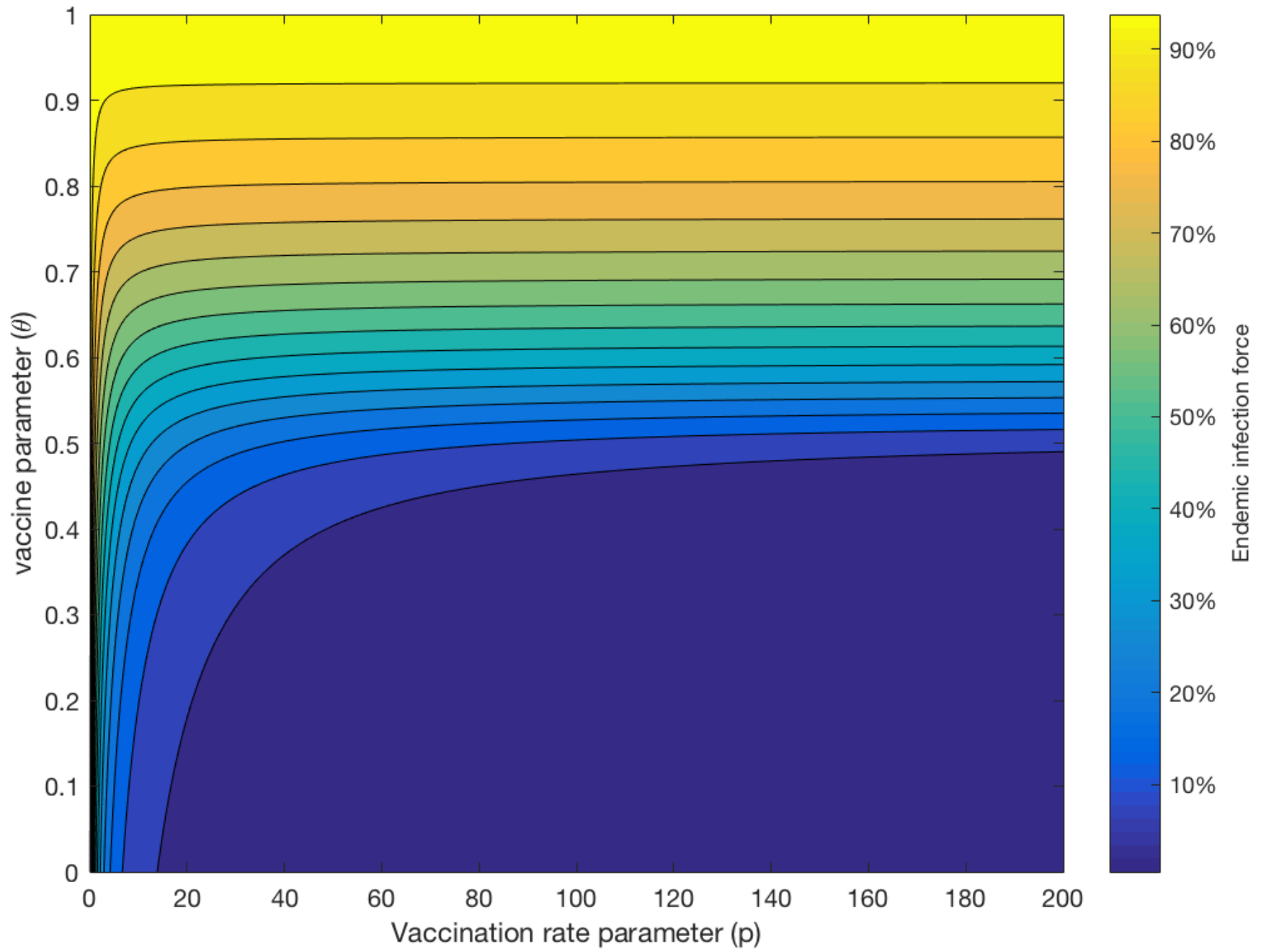}   
 \caption{\tiny  Cumulative force of infection $y(\infty)$ in percentage (vaccination versus non vaccination scenario)} 
 \vspace{-10pt}
\label{fig:sub13}
\end{subfigure}
\caption{Impact of adaptive vaccination strategy on cumulative force of infection $y(\infty)$.}
\label{fig:test1}
 \vspace{-10pt}
\end{figure}

 \noindent
\textbf{Case 4: $0<\theta<1$ and $p>0$}\\
The expression in  \eqref{eq:8} can be transformed into the following  standard equation \eqref{eq:12}  with one unknown variable $x$, which is the variable of interest.
 \begin{equation} 
\label{eq:12}
\begin{split}
x-R\left(1-\frac{pe^{-\theta x}+(1-\theta)e^{-(1+p)x}}{1+p-\theta}\right) \;=\; 0 \hspace{7mm}
\hbox{where}  \ \
 x=y(\infty) 
 \end{split}
\end{equation}
 \noindent
To consider a representative situation, we supposed the reproduction number ($R$) is equal to 2. As illustrated  in the Figure \ref{fig:test1}, the result shows that the effective vaccine is the most determinant  factor.  In fact, when the vaccine is not effective (high vaccine parameter), whatever the vaccination rate parameter chosen, the effect on $y(\infty)$ is marginal (yellow area). On the other side, $y(\infty)$ is responsive to both factors when the vaccine parameter is low; the degree of $y(\infty)$ reduction depends on the vaccination rate parameter. As shown in Figure \ref{fig:test1}, the color on the graph becomes blue quickly when the vaccination rate parameter increases. In Figure \ref{fig:test1}, the graphs \ref{fig:sub11} and \ref{fig:sub12} are two views of the same function in a three-dimensional space, whereas the graph \ref{fig:sub13} is a two-dimensional space. 

\newpage
\section{ Age-structured epidemic model}
\noindent
 In this section, we consider the situation where, at the population level, new susceptibles arise as a result of reproduction at a constant birth rate $B$. In addition, we consider the survival function $\scrF(a)$, which describes the probability that a newborn individual lives at least until age $a$. If at time $t$ a susceptible has age $a$, then at time $t-a+\sigma$  this susceptible has age $\sigma$ 
($0<\sigma\leq a$). For a small positive time duration $h$, taking into account the survival function $\scrF(a)$, the behavior of the non-vaccinated susceptibles $S(t,a)$ at time $t$ and  at age $a$ follows the equation 
\begin{equation}
  \label{eq:13}
S(t-a+\sigma+h,\sigma+h)\;=\;S(t-a+\sigma,\sigma)\frac{\scrF(\sigma+h)}{\scrF(\sigma)}
  (1-F(t-a+\sigma)h-\phi(t-a+\sigma)h+o(h^2)) \,.
\end{equation}
 By re-arranging, we have the following approximation.
\begin{equation*} 
\begin{split}
\frac{\frac{S(t-a+\sigma+h,\sigma+h)}{\scrF(\sigma+h)} - \frac{S(t-a+\sigma,\sigma)}{\scrF(\sigma)}}{h}=-F(t-a+\sigma)\frac{S(t-a+\sigma,\sigma)}{\scrF(\sigma)}-\phi(t-a+\sigma)\frac{S(t-a+\sigma,\sigma)}{\scrF(\sigma)}+\frac{o(h^2)}{h}.
\end{split}
\end{equation*}
We can use similar reasoning for the vaccinated susceptibles $V(t,a)$ at time $t$ and  at age $a$, and
then take the limit as $h$ converges to 0,  resulting in the system of differential equations  
\begin{equation} 
\label{eq:16}
\begin{split}
\frac{d(\frac{S(t-a+\sigma,\sigma)}{\scrF (\sigma)})}{d\sigma} &\;=\;
-F(t-a+\sigma)\frac{S(t-a+\sigma,\sigma)}{\scrF(\sigma)}\;-\;\phi(t-a+\sigma)\frac{S(t-a+\sigma,\sigma)}{\scrF(\sigma)}, \\
\frac{d(\frac{V(t-a+\sigma,\sigma)}{\scrF (\sigma)})}{d\sigma}  &\;=\;
-\theta F(t-a+\sigma)\frac{V(t-a+\sigma,\sigma)}{\scrF(\sigma)}\;+\;\phi(t-a+\sigma)\frac{S(t-a+\sigma,\sigma)}{\scrF(\sigma)}.
\end{split}
\end{equation}
\noindent
 If there is no infection in the population, then we have a stable age  distribution (Breda et al (2012)), with
$S(t,a)$ and $V(t,a)$ becoming
\begin{equation*} 
\begin{split}
S(t,a)&\;=\;B\scrF (a) \,, \\
V(t,a)&\;=\;0 \,.
\end{split}
\end{equation*} 
\noindent
Recall from the transfer diagram of the model that $B$  is the constant birth rate.\\
More generally, by solving the system of equations \eqref{eq:16}, we obtained the  following results.
\begin{equation} 
\label{eq:16a}
\begin{split}
S(t,a)&=B\scrF (a)e^{-\int_{0}^{a}(F+\phi)(t-a+\sigma)d\sigma}, \\
V(t,a)&=B\scrF (a)e^{-\theta\int_{0}^{a}F(t-a+\sigma)d\sigma} \int_{0}^{a}\phi(t-a+\sigma)e^{-\int_{0}^{\sigma}((1-\theta)F+\phi)(t-a+\tau)d\tau} d\sigma.
\end{split}
\end{equation}
\noindent
The force of infection depends  on   the size of the infectious population and the survival function characteristics of the population. At time $t$, from individuals who were infected at time $t-\tau$  at age $a$, the contribution to the force of infection is the product of $(F(t-\tau)S(t-\tau,a)+ \theta F(t-\tau)V(t-\tau,a))A(\tau)$ infectious individuals  and a demographic factor $\frac{\scrF (a+\tau)}{\scrF (a)}$, which is the fraction of infectious individuals  who survive from age $a$ to age $a+\tau$. By summing all the contributions with respect to to the elapsed time $\tau$ and with respect to the age $a$, we get the following renewal equation.
\begin{equation} 
\label{eq:17}
\begin{split}
F(t) &\;=\; \int_{0}^{\infty}\int_{0}^{\infty} (F(t-\tau)S(t-\tau,a)+ \theta F(t-\tau)V(t-\tau,a))A(\tau)\frac{\scrF (a+\tau)}{\scrF (a)} \ d\tau \ da  \\
     &\;=\; \int_{0}^{\infty}F(t-\tau) \int_{0}^{\infty} (S(t-\tau,a)+ \theta V(t-\tau,a))\frac{\scrF (a+\tau)}{\scrF (a)}A(\tau) \ d\tau \ da \,.
\end{split}
\end{equation}
\noindent
The integral  $\int_{0}^{a}\phi(t-a+\sigma)e^{-\int_{0}^{\sigma}((1-\theta)F+\phi)(t-a+\tau)d\tau} d\sigma$ from \eqref{eq:16a} is not easy to handle  because the rate of vaccination function $(\phi(t))$ is unknown.
As in Section 2, we assume a linear relationship between the  vaccination rate and the force of infection, in
the form $\phi(t)\,=\,pF(t)$  where $p$ is the vaccination rate parameter. The solution \eqref{eq:16a} becomes:
\begin{equation} 
\label{eq:19}
\begin{split}
S(t,a)&=B\scrF (a)e^{-(1+p)\int_{0}^{a}F(t-a+\sigma)d\sigma}  \,,  \\
V(t,a)&=B\scrF (a)\frac{p}{1+p-\theta}\left(e^{-\theta\int_{0}^{a}F(t-a+\sigma)d\sigma} -e^{-(1+p)\int_{0}^{a}F(t-a+\sigma)d\sigma}\right) \,.
\end{split}
\end{equation}

\subsection{Characteristic equation of the endemic steady state}
\noindent
The solution \eqref{eq:19} can be substituted into the renewal equation \eqref{eq:17}. In the endemic steady state, we assume the force of infection ($F(t)$) converges to a constant value $F$. When $t$ goes to $+\infty$, the renewal equation can be rearranged, leading to the following characteristic equation of the endemic steady state.
\begin{equation} 
\label{eq:20}
 1\;=\;\frac{B}{1+p-\theta}\int_{0}^{\infty}\int_{0}^{\infty} \left(p\theta e^{-\theta aF}+(1+p)(1-\theta)e^{-(1+p)aF}
    \right) \scrF (a+\tau)\,A(\tau) \, d\tau \, da \,.
\end{equation}
\noindent
In order to study the properties of the characteristic equation, we consider the following function:
\begin{equation} 
\label{eq:21}
 f(x)\;=\;\frac{B}{1+p-\theta}\int_{0}^{\infty}\int_{0}^{\infty} \left(p\theta e^{-\theta ax}+(1+p)(1-\theta)e^{-(1+p)ax}
 \right) \scrF (a+\tau)\,A(\tau) \, d\tau \, da \,.
\end{equation}
\noindent
It is obvious that $f(x)$ is decreasing in $x$,   and 
$f(0)=B\int_{0}^{\infty}\int_{0}^{\infty}\scrF (a+\tau)A(\tau) \, d\tau \, da$. The condition $f(0)>1$ is sufficient 
to guarantee the existence of a solution $F$ of equation \eqref{eq:20} with $F \neq 0$. 
Furthermore, $f(0)$ depends only on the constant birth rate $B$, the survival function $\scrF$, 
and the expected contribution to the force of infection $A$, since
the vaccine parameters $\theta$ and $p$ cancel out.  Indeed, as noted by Breda et al (2012), 
$f(0)$ is  the basic reproduction number and can interpreted as the expected number of secondary cases caused by a primary case introduced in a susceptible population with age distribution $B\scrF(a)$.
\begin{equation} 
\label{eq:22}
 R\;=\;f(0)\;=\; B\int_{0}^{\infty}\int_{0}^{\infty}\scrF (a+\tau)\,A(\tau) \,d\tau \, da \,.
\end{equation}

\subsection{General case}
\noindent
Throughout this subsection, we shall assume $R \,=\, f(0) \, >\,1$.
\noindent
The implicit function theorem can be used on equation \eqref{eq:20}  to express $F$ as a function of two variables $p$ and $\theta$. \\
Let us define
$$G(\theta,p,F)=\frac{B}{1+p-\theta}\int_{0}^{\infty}\int_{0}^{\infty} (p\theta e^{-\theta aF}+(1+p)(1-\theta)e^{-(1+p)aF})\scrF (a+\tau)A(\tau) \ d\tau \ da -1.$$
Then $G$ is a continuously differentiable function;  $G(\theta,p,F)=0$ if $F$ is the solution of equation \eqref{eq:20}; and $\frac{dG(\theta,p,F)}{dF}<0$.
Therefore $F(\theta,p)$ is a continuously differentiable function and the solution of the following  equation.
\begin{equation} 
\label{eq:31}
 1=\frac{B}{1+p-\theta}\int_{0}^{\infty}\int_{0}^{\infty} (p\theta e^{-\theta aF(\theta,p)}+(1+p)(1-\theta)e^{-(1+p)aF(\theta,p)})\scrF (a+\tau)A(\tau) \ d\tau \ da .\\
\end{equation}
\smallskip

\noindent
\textbf{Case 1: ineffective vaccine ($\theta =1$)}\\
When $\theta=1$, the equation \eqref{eq:31} becomes
\begin{equation} 
\label{eq:32}
 1=B\int_{0}^{\infty}\int_{0}^{\infty} e^{-aF(1,p)}\scrF (a+\tau)A(\tau) \ d\tau \ da .
\end{equation}
\noindent
We have $F(1, p)=F^*$, where $F^*$ comes from the non-vaccination case studied by Breda et al (2012). Therefore,  the endemic force of infection is  the same as the endemic force of infection without vaccination.\\

\noindent
\textbf{Case 2: 100\% effective vaccine ($\theta= 0$)} \\
When $\theta=0$, the equation \eqref{eq:31} becomes
 \begin{equation} 
\label{eq:33}
 1=B\int_{0}^{\infty}\int_{0}^{\infty} e^{-(1+p)aF(0,p)}\scrF (a+\tau)A(\tau) \ d\tau \ da .
\end{equation}
\noindent
Similarly to Case 1, we have $(1+p)F(0,p) =F^*$ for  $p>0$.
Thus  $F(0,p) =\frac{F^*}{p+1}$, which depends on the factor  $\frac{1}{p+1}$, and we see that
$\lim_{p\to\infty}F(0,p)=0$, which corresponds to the disease-free steady state.\\

\noindent
\textbf{Case 3: $p \rightarrow \infty$ and $\theta\neq0$}\\
We assumed  $\lim_{p\to\infty}F(\theta,p)=F(\theta, +\infty)$.\\
We shall show that the solution of the equation \eqref{eq:31} becomes:
\begin{equation}
   \label{eq:30A}
    F(\theta,+\infty)=
  \begin{cases}
  0     &  \text{if } \theta\leq \frac{1}{f(0)} \,, \\
   F(\theta,+\infty) &  \text{if } \frac{1}{f(0)}< \theta<1 \,,\\
 F^*   &  \text{if } \theta=1  \,. 
  \end{cases}
\end{equation}
In addition, for $\frac{1}{f(0)}< \theta<1$, $F(\theta,+\infty)$ is an  increasing function and $0< F(\theta,+\infty)<F^*$.\\

\noindent
a) For $\theta=1$ : \\
 By Case 1 above, $F(1, p)=F^* $ for all $p>0$. Therefore,  $F(1,+\infty)=F^*$.
 \\
\noindent 
\\
b) For $0<\theta\leq \frac{1}{f(0)}$ :  \\
 We shall assume $F(\theta,+\infty)>0$.\\
Let us take a sequence $(p_n)$ such that $\lim_{n \to \infty}p_n=+\infty$,
so that $\lim_{n \to \infty}F(\theta,p_n)=F(\theta, +\infty)$. \\
 We consider  the sequence of functions $(f_n)$ defined by
\begin{equation*} 
f_n(a,\tau)=\frac{B}{1+p_n-\theta} \left(p_n\theta e^{-\theta aF(\theta,p_n)}+(1+p_n)(1-\theta)e^{-(1+p_n)aF(\theta,p_n)}\right)\scrF (a+\tau)A(\tau) \,.
\end{equation*}
According to the definition of $F(\theta,p_n)$ and the equation  \eqref{eq:31}, we have $\int_{0}^{\infty}\int_{0}^{\infty}f_n(a,\tau) \ d\tau\ da=1$.
The function $f_n$ is dominated by an integrable function $ B\scrF (a+\tau)A(\tau)$. In fact, we have
 \begin{equation*} 
f_n(a,\tau) \;\leq \; B\,\scrF (a+\tau)A(\tau)
     \hspace{5mm} \hbox{\and}  \hspace{5mm}
f(0) \;=\; \int_{0}^{\infty}\int_{0}^{\infty}B\scrF (a+\tau)A(\tau)\, d \tau\, da \,.
\end{equation*}
 \noindent
Following the Dominated Convergence Theorem (DCT), and recalling $F(\theta,+\infty)>0$, we have
 \begin{equation} 
  \label{eq:30B}
   1\;=\; \int_{0}^{\infty}\int_{0}^{\infty}\lim_{n \to \infty}f_n(a,\tau)\,  da\  d\tau \;=\;
 B\theta\int_{0}^{\infty}\int_{0}^{\infty}e^{-\theta aF(\theta, +\infty)}\scrF (a+\tau)A(\tau) \ d\tau \ da 
\end{equation}
and we have 
 \begin{equation*} 
   1\;=\;B\theta\int_{0}^{\infty}\int_{0}^{\infty}e^{-\theta aF(\theta, +\infty)}\scrF (a+\tau)A(\tau) \ d\tau \ da \;<\; \theta f(0) .
\end{equation*}
This means $1<\theta f(0)$, which is a contradiction. 
Therefore, $F(\theta,+\infty)= 0$ for $0<\theta\leq\frac{1}{f(0)}$.\\
\noindent 
\\
c) For $\frac{1}{f(0)}< \theta<1$ :  \\ 
Suppose  $F(\theta,+\infty)=0$.\\
From $\eqref{eq:31}$ 
 \begin{equation*} 
   1\;\geq \;B\frac{p\theta}{1+p-\theta}\int_{0}^{\infty}\int_{0}^{\infty} e^{-\theta aF(\theta,p)}\scrF (a+\tau)A(\tau) \ d\tau \ da .
   \end{equation*}
When $p$ goes to $+\infty$,
 \begin{equation*} 
1\;\geq \; \theta B\int_{0}^{\infty}\int_{0}^{\infty} e^{-\theta aF(\theta,+\infty)}\scrF (a+\tau)A(\tau) \ d\tau \ da 
   \;=\; \theta f(0) \,,
   \end{equation*}
which means  $\theta \leq \frac{1}{f(0)}$ ; and this is a contradiction. Therefore, $F(\theta,+\infty) \neq 0$. \\
\noindent 
\\
d) Finally, we show that $F(\theta,+\infty)$ is an  increasing function on the interval $\left(\frac{1}{f(0)},1\right)$.\\
Let us consider a function $\psi_\theta (x)$ defined for every $x \geq 0$ and $ \theta \geq 0$ by
\begin{equation*} 
 \psi_\theta (x)\;=\; B\theta\int_{0}^{\infty}\int_{0}^{\infty} e^{-\theta ax}\scrF (a+\tau)A(\tau) \ d\tau \ da .
\end{equation*}
$ \psi_\theta (x)$ is well defined because $ \psi_\theta (x)\leq f(0)$.
It can be checked that $ \psi_\theta (x)$ is a strictly decreasing continuous function of $x$.
According to the Dominated Convergence Theorem (DCT)  in  $\eqref{eq:30B}$,
\begin{equation*} 
\psi_\theta (F(\theta, +\infty))\;=\;1\,.
\end{equation*}
By changing the variable,  $\psi_\theta (x)$ becomes 
\begin{equation*} 
 \psi_\theta (x)\;=\;B\int_{0}^{\infty}\int_{0}^{\infty} e^{-bx}\scrF \left(\frac{b}{\theta}+\tau\right)A(\tau) \ d\tau \ db \,,
\end{equation*}
which  shows that $ \psi_\theta (x)$ is an increasing function of $\theta$.
Let us take $\theta_1$ and $\theta_2$ such that $\frac{1}{f(0)}<\theta_1<\theta_2<1$.
Then we have:
  \begin{equation}
\label{eq:36a}
  \psi_{\theta_2} (F(\theta_1, +\infty))\;\geq \; \psi_{\theta_1} (F(\theta_1, +\infty))\;=\;1\;=\; \psi_{\theta_2} (F(\theta_2, +\infty)) .\\     \end{equation}
 \noindent 
  $\psi_\theta (x)$ is a strictly decreasing function of $x$ and from  the development \eqref{eq:36a}, we conclude that  $F(\theta_1, +\infty)\leq F(\theta_2, +\infty)$. This shows that $F(\theta, +\infty)$ is an  increasing function.\\

\subsection{Special case of natural constant per-capita mortality rate ($\mu$)}
\noindent
In this subsection, we assume that all individuals have a survival function $\scrF (a)=e^{-\mu a}$, which describes a constant per-capita mortality rate $\mu$.
\noindent
By applying the survival function, the following basic reproduction number is derived from \eqref{eq:22}:
\begin{equation} 
\label{eq:23}
\begin{split}
R\;=\; f(0) \; &=\; B\int_{0}^{\infty}\int_{0}^{\infty}e^{-\mu (a+\tau)} A(\tau) \ d\tau \ da \\
      &= \;\frac{B}{\mu}\int_{0}^{\infty}e^{-z \tau}A(\tau) d\tau \, .
 \end{split}
\end{equation}
 \noindent
 This expression for the reproduction number was also found by Breda et al (2012) for a constant per-capita mortality rate $\mu$. The characteristic equation \eqref{eq:20} for endemic steady state becomes  a second degree equation. The coefficients depend on the parameters of vaccination and the reproductive number.
\begin{equation} 
\label{eq:24}
\frac{\theta(1+p)}{\mu f(0)}F(\theta,p)^2 +\left(\frac{1+p+\theta}{f(0)}-\theta (1+p)\right) F(\theta,p) + \frac{\mu}{f(0)}(1-f(0)) \;=\;0 .
\end{equation}
\noindent
We define:
\begin{align*}
a & \;=\; \frac{\theta(1+p)}{\mu f(0)}  ,  \\
b & \;=\;\frac{1+p+\theta}{f(0)}-\theta (1+p) ,  \\ 
 c &\; =\; \frac{\mu}{f(0)}(1-f(0)) .
\end{align*}
\noindent
In an endemic steady state, $f(0)>1$, leading to $c=\frac{\mu}{f(0)}(1-f(0))<0$ and $b^2 - 4ac>0$.
\noindent
The solution of the equation \eqref{eq:24} becomes
\begin{equation} 
\label{eq:26}
\begin{split}
F(\theta,p) & \;=\; \frac{-b+ \sqrt{b^2 - 4ac}}{2a} \\
   & \;=\;\frac{\mu}{2}\left\{\left(f(0)-\frac{1+p+\theta}{\theta (1+p)}\right)+ \sqrt{\left(\frac{1+p+\theta}{\theta (1+p)}-f(0)\right)^2 + 4\frac{f(0)-1}{\theta (1+p)}} \right\} .
\end{split}
\end{equation}
\noindent
The endemic force of infection is more complex  with vaccination parameters. In fact, in Breda et al (2012), with a constant per-capita mortality rate $\mu$ and no vaccination, the authors find the endemic force of infection  $F^* =\mu(f(0)-1)$.\\

\noindent
\textbf{Case 1: ineffective vaccine ($\theta =1$) }

 The endemic force of infection becomes
 \begin{equation*} 
\begin{split}
F(1,p)\;=\; \lim_{\theta \to 1}F(\theta,p) \;& =\; \frac{\mu}{2}\left\{\left(f(0)-\frac{2+p}{1+p}\right)+ \sqrt{\left(\frac{2+p}{1+p}-f(0)\right)^2 + 4\,\frac{f(0)-1}{1+p}}\right\} \\  
 & =\; \frac{\mu}{2}\left\{\left\{f(0)-1- \frac{1}{1+p}\right)+ \sqrt{\left(f(0)-1+ \frac{1}{1+p}\right)^2}\right\}   \\
  & =\; \mu (f(0)- 1) \\
  & =\; F^* .
    \end{split}
\end{equation*}
\noindent
In the case of 100\% ineffective vaccine, the endemic force of infection is  the same as the endemic force of infection without vaccination (Breda et al (2012)).\\

\noindent
\textbf{Case 2: 100\% effective vaccine ($\theta= 0$)}  \\
Multiplying the quadratic formula of $\eqref{eq:26}$ by $b+\sqrt{b^2-4ac}$ in numerator and denominator gives
\[
F(\theta,p) \;=\;    \frac{-b^2 +(\sqrt{b^2-4ac})^2}{2a (b+\sqrt{b^2-4ac})}   
      \;=\;    \frac{-2c}{b+\sqrt{b^2-4ac}}   \,.
\]
As $\theta\rightarrow 0$, we see that $a\to 0$ and $b\to (1+p)/f(0)$, resulting in 
\[    \lim_{\theta\to 0} F(\theta,p)    \;=\;   \frac{-2\mu(1-f(0))/f(0)}{2(1+p)/f(0)}   \;=\;   
      \frac{\mu(f(0)-1)}{1+p}    \;=\;   \frac{F^*}{1+p}   \,.
\]
\noindent
Again, $F^* =\mu(f(0)-1)$ is the endemic force of infection without vaccination from Breda et al (2012).
Taking into account the vaccination process, we have  $F =\frac{F^*}{p+1}$, which depends on the factor  
$\frac{1}{p+1}$, with vaccination rate parameter $p$. We have $\lim_{p \to \infty}\left\{\lim_{\theta \to 0}F(\theta,p)\right\} = 0$, which corresponds to the disease-free steady state.\\

\noindent
\textbf{Case 3: $p \rightarrow \infty$ and $\theta\neq0$}\\
The vaccine is not 100\% effective. By increasing the vaccination rate parameter ($p$), the expression in \eqref{eq:26} becomes
 \begin{equation*} 
\begin{split}
\lim_{p \to \infty}F(\theta,p) \;& =\; \lim_{p\to\infty}\frac{\mu}{2}\left\{(f(0)-\frac{1+p+\theta}{\theta (1+p)})+ \sqrt{\left(\frac{1+p+\theta}{\theta (1+p)}-f(0)\right)^2 + 4\frac{f(0)-1}{\theta (1+p)}} \right\} \\
   & =\; \frac{\mu}{2}\left\{(f(0)-\frac{1}{\theta })+ \sqrt{\left(\frac{1}{\theta }-f(0)\right)^2 } \right\} .\\
\end{split}
\end{equation*}
\[ \therefore  \hspace{5mm}\lim_{p \to \infty}F(\theta,p)=
  \begin{cases}
  0     & \quad \text{if } \theta\leq \frac{1}{f(0)} \,, \\
 \mu(f(0)-\frac{1}{\theta })   & \quad \text{if } \frac{1}{f(0)}< \theta\leq 1\,.
  \end{cases}
\] 
With a sufficiently high vaccination rate parameter ($p$), the disease-free steady state can still be approached
arbitrarily closely if the vaccine parameter ($\theta$) is below a threshold $\left(\theta\leq\frac{1}{f(0)}\right)$. \\

\begin{figure}[h]  
    \begin{subfigure}{.47\linewidth}
\centering
    \includegraphics[scale=0.4]{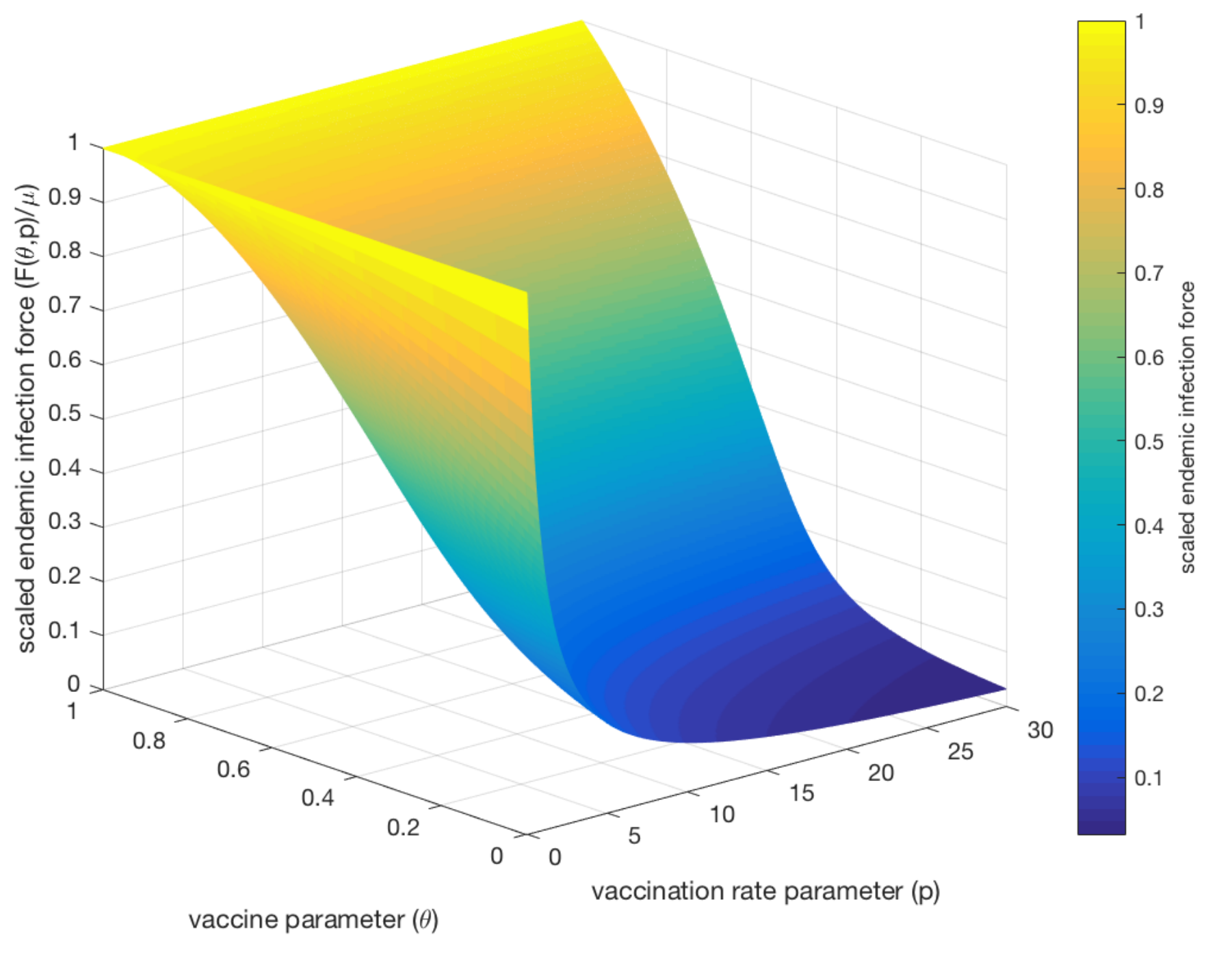}  
\caption{\tiny Endemic force of infection as a function of $\theta$ and $p$}
\vspace{10pt}
\label{fig:1a}
\end{subfigure}%
\begin{subfigure}{.5\linewidth}
\centering
    \includegraphics[scale=0.4]{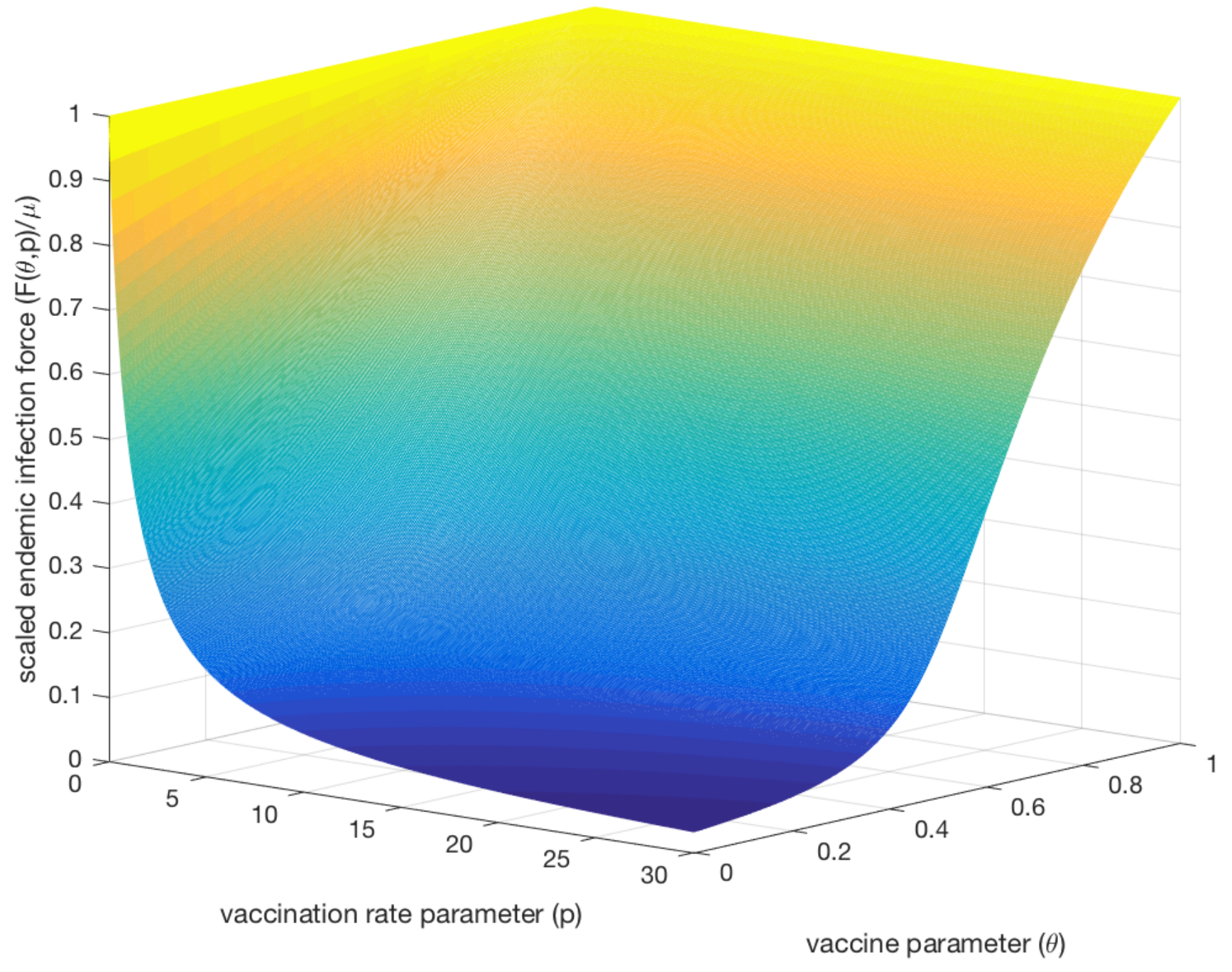}   
\caption{\tiny Endemic force of infection as a function of $\theta$ and $p$}
  \vspace{10pt}
\label{fig:1b}
\end{subfigure}\\[1ex]
\begin{subfigure}{\linewidth}
\centering
 \vspace{-15pt}
    \includegraphics[scale=0.5]{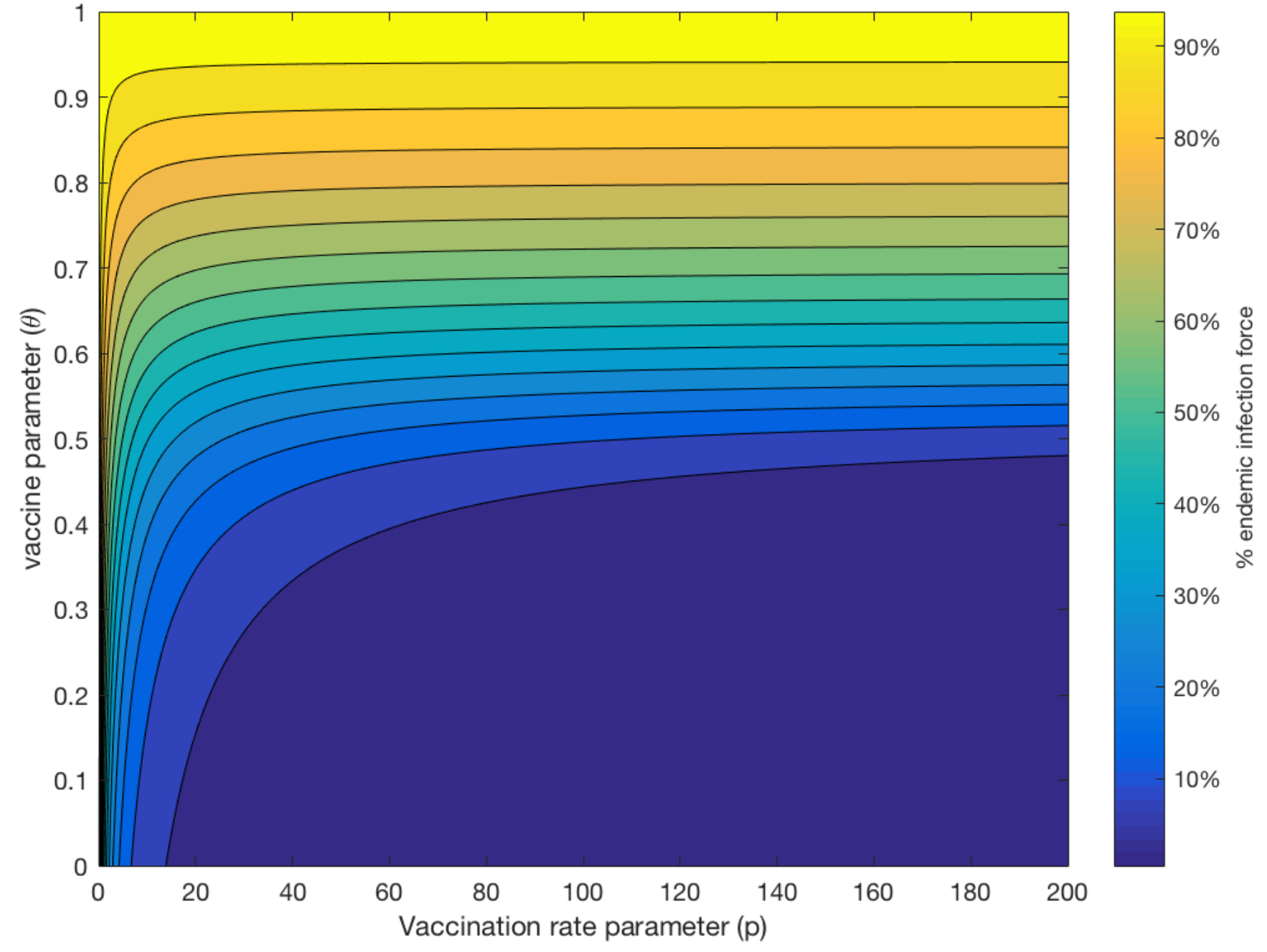}   
 \caption{\tiny Endemic force of infection in percentage (vaccination versus non-vaccination scenario)} 
 \vspace{-10pt}
\label{fig:1c}
\end{subfigure}
\caption{ Impact of adaptive vaccination strategy on the endemic  force of infection ($R=f(0)=2$).}
\label{fig:test2}
 \vspace{-20pt}
\end{figure}

\noindent
\textbf{Case 4: $0<\theta<1$ and $p>0$} \\
The endemic force of infection in the expression \eqref{eq:26} was simulated as a function of vaccination rate parameter ($p$) and vaccine parameter ($\theta$). Figure \ref{fig:test2} below  provides a summary of the findings. We supposed the reproduction number ($R$) is equal to $2$. 
As illustrated by the yellow area in Figure \ref{fig:test2}, the scaled endemic infection force 
$\left(\frac{F^*}{\mu}\right)$ remains almost constant; whereas in the blue area,  the reduction of  the endemic force of infection is significant. Compared to the non-vaccination scenario,  the endemic force of infection is almost 0. In Figure \ref{fig:test2}, the graphs \ref{fig:1a} and \ref{fig:1b} are two views of the same function in a three-dimensional space, whereas the graph \ref{fig:1c} is a two-dimensional space.

\section{Conclusion}
\noindent
The force  of infection  was developed by Breda et al (2012) to study the disease spread  within a closed  population  and a demographically open population. In the same context of disease spread, the susceptibles in each population are divided into two groups, non-vaccinated susceptible and vaccinated susceptible,  in order to evaluate the combined effect of vaccine effectiveness and the vaccination rate on the dynamics of the force of infection. The adaptive vaccination strategy  is used  and consists of choosing the rate of vaccination proportional to the force of infection; although simplistic,  this natural strategy makes the model more analytically tractable. 
 The investigation focuses on the relation between force of infection of  the disease spreading, vaccine effectiveness, and  adaptive vaccination rate. The vaccine parameter $\theta$ ($0\leq \theta \leq 1$) captures the vaccine effectiveness, and the vaccination rate parameter ($p>0$) determines the rate of vaccination. 
\noindent
 As shown by the results, the Reproduction Number ($R$) can be used to compare the model  without vaccination (Breda et al (2012)) and with vaccination. In fact in each population, $R$ is not affected by  the vaccine parameter and  the vaccination rate parameter. The findings show that the cumulative force of infection in a closed population and the endemic force of infection in a demographically open population have the same pattern of behavior as  they can be reduced significantly  by the same two factors: the effective vaccine  and the vaccination rate. In fact, for a given vaccination rate parameter ($p$), the cumulative force of infection or the endemic force of infection decreases  significantly when the vaccine is effective ($\theta$ below a threshold (dark blue color in Figure \ref{fig:test1} and Figure \ref{fig:test2})). When $p$ is sufficiently big, the threshold of effectiveness is $\frac{1}{R}$ ($\theta<\frac{1}{R}$) for  respectively closed population and  demographically open population; and the force of infection transforms an endemic steady state into to a disease-free state. The study also highlights  the fact  that only the effective vaccine is the key for  a successful reduction of the force of infection. In fact, when the vaccine is not effective ($\frac{1}{R}<\theta$), for any vaccination rate parameter  chosen, the effect on the cumulative force of infection ($y(\infty)$) or the endemic force of infection 
 $\left(\frac{F^*}{\mu}\right)$ remains significantly different from 0.  The results also show that $\theta R$  serves as an effective Reproduction Number when $p$ is large. 
 
 \section*{acknowledgements}
 \noindent
 This research was supported in part by a Discovery Grant to N. Madras from the Natural Sciences and Engineering Research Council of Canada. The authors wish to thank Mahnaz Alavinejad and Jianhong Wu for discussions of the paper of Breda et al (2012).\\

\end{document}